\documentstyle[12pt]{article}
\begin{document}
\setcounter{footnote}{0}
\renewcommand{\thefootnote}{\fnsymbol{footnote}}
\begin{flushright}
NSF-ITP-95-67\\
CTP-TAMU-14/95\\
NUB-TH-3121/95\\
\end{flushright}
\begin{center}
{\bf Predictions of Neutralino Dark Matter Event Rates in Minimal
Supergravity Unification}\\
\vspace{1.5cm}
{\bf R. Arnowitt}\\
Center for Theoretical Physics, Department of Physics\\
 Texas A\&M University, College Station, TX  77843-4242\\
\vspace{1cm}
{\bf Pran Nath}\\
\footnote {Permanent address}Department of Physics, Northeastern
University\\ Boston, MA  02115 \\
\& \\
Institute for Theoretical Physics\\
University of California, Santa Barbara, CA 93106\\
\end{center}
\begin{abstract}
A detailed analysis of dark matter event rates in minimal supergravity models
(MSGM) is given.  It is shown analytically that the lightest neutralino the
${{\tilde {Z}_{1}}}$ is the LSP over almost all of the parameter space,
and hence
the natural candidate for cold dark matter (CDM).  The radiative breaking of
$SU(2)\times U(1)$ constraints are shown to be crucial in determining the
expected event rates.  Approximate analytic formulae are obtained to determine
the gaugino-higgsino content of the ${{\tilde{Z}_{1}}}$ particle.  From
this one
can deduce the behavior of the event rates as one varies the SUSY soft
breaking parameters and tan $\beta$.  The constraint on the event rates due to
the recently measured $b\rightarrow s+\gamma$ decay is calculated.  It is
seen that this data eliminates most of the parameter space where $\mu$ (the
Higgs mixing parameter) and $A_t$ (the t-quark cubic soft breaking
parameter) have the same sign.  Since the t-quark is close to its Landau
pole, $A_t$ is restricted to be mostly positive, and so most of the
$\mu>0$ part of the parameter space is eliminated.  However, for $\mu<0$,
one finds large regions of parameter space where the event rate is large and
exceeds 0.01 events/kg da.  The importance of proper treatment of the
s-channel Z and Higgs poles in calculating the relic density is stressed.
The implications of the recent new experiments (SMC and E143) on the quark
polarizabilities are analysed and it is seen that uncertainties in these
generally produce only  small uncertainties in the event rates.  A discussion
is also given of the sensitivity of the expected event rates to changes in
the allowed range of ${{\tilde {Z}_{1}}}$ relic density.
\end{abstract}

\newpage
\noindent
{\bf 1.~~Introduction}\\
{}~\\
\indent
The nature of the dark matter (DM) which makes up more than 90\% of the matter
of the universe is a particularly important issue as it may have a fundamental
impact both on astronomy and particle physics.  Dark matter has currently only
been detected by its gravitational interactions, and thus may be composed of
several constituents, e.g. baryonic dark matter (B), hot dark matter
(HDM) and cold dark matter (CDM) (where ``hot'' and ``cold'' refer to whether
the particle was relativistic or non-relativistic at the time of galaxy
formation).  One may measure the amount of each species of dark matter by the
ratio $\Omega_{i}=\rho_{i}/\rho_{c}$ where $\rho_{i}$ is the mass density of
the
ith constituent and $\rho_{c}=3H^{2}/8 \pi G_{N}$ is the critical mass density
(H is the Hubble constant and $G_{N}$ is the Newtonian gravitational
constant).  Within the framework of the inflationary scenario one has
$\Sigma\Omega_{i}=1$.  The amount of baryonic matter is severely limited in the
big bang cosmology by the observed abundancies of light elements i.e.
$\Omega_{B}\stackrel{<}{\sim}$ 0.1.\\
\indent
Rotation curves of stars imply a density of dark matter in our galaxy of
\begin{equation}
\rho_{DM}\cong 0.3GeV/cm^{3}
\end{equation}
\noindent
and this matter will be impinging on the Earth with velocity $v_{DM}\simeq
320$ km/s.
The fact that microlensing find far more machos in the disk than in the halo of
the Galaxy implies that at most 30\% of the halo dark matter is machos [1].
Thus most of the halo of the Galaxy must be cold dark matter, and it is this
dark matter that terrestial detectors can observe.

A possible source of hot dark matter is massive neutrinos.  In this paper we
assume that the cold dark matter component is the lightest supersymmetry
neutralino, the $\tilde{Z}_{1}$ particle.  The anisotropy power spectrum
(including the recent COBE data) puts contraints on the relative amounts of HDM
and CDM.  A reasonable fit to the full spectrum gives
$\Omega_{CDM}:\Omega_{HDM}=2:1$.  Assuming $\Omega_{B}\simeq 0.1$ one then
estimates $\Omega_{\tilde{Z}_{1}}=0.6$.  What is theoretically calculable is
$\Omega{h}^{2}$ where h=(H/100 km/s Mpc).  Current astronomical measurements
yield \begin{equation}
0.4\stackrel{<}{\sim}h\stackrel{<}{\sim}0.8
\end{equation}
\noindent
i.e. two groups of measurements of h exist, one clustering at the lower bound
and one at the upper bound.  [The inflationary scenario (with zero
cosmological constant) requires $h\simeq 0.5$]. Thus we estimate
\begin{equation}
0.1\leq\Omega_{\tilde{Z}_{1}}h^{2}\leq 0.4
\end{equation}
\noindent
and we will assume these bounds in the following.  (Our results are not
qualitatively
sensitive to the precise upper and lower limits of Eq. (3) and we will discuss
below what changes occur if one perturbs them.)   We also note that it has
recently been suggested [2] that if the value $h\simeq 0.8$ is correct, the
age of the universe and other cosmological problems could be accounted for
by a cosmological constant with $\Omega\simeq 0.6-0.8$, and the remainder
being CDM.  This would also lead to $\Omega_{\tilde{Z}_{1}}$$h^2$ being in the
range of Eq. (3).

Current dark matter detectors plan to obtain a sensitivity of $R\simeq 0.1$
events/kg~da.  Future developments may improve this to R=0.01 events/kg da.  We
will thus limit our discussion here to the part of the parameter space where
\begin{equation}
R\stackrel{>}{\sim} 0.01~events/kg da
\end{equation}
\noindent
since this sensitivity is what one may expect over the next 5-10 years.
Detection of the $\tilde{Z}_{1}$ depends on their scattering cross section by
quarks in the nuclei of the detector.  Thus a calculation of event rates
depends on the two things:~~(i) that the relic density of $\tilde{Z}_{1}$ obey
the bounds of Eq. (3), which limits the allowed SUSY parameter space, and (ii)
a calculation of the  $\tilde{Z}_{1}-q$ cross section.  We consider these
calculations in this paper within the framework of supergravity grand
unification
models [3].  While this model is not a complete theory it possesses a
sufficient number of accomplishments to warrent using it as the dynamical
framework.  Thus it accounts naturally for grand unification at a scale
$M_{G}\simeq 10^{16}$ GeV implied by the LEP measurements of $\alpha_{1},
\alpha_{2}$ and $\alpha_{3}$ at $M_{Z}$; it allows a natural breaking of
supersymmetry (in the hidden sector) at the GUT scale (something that cannot be
done in a phenomenologically acceptable way in the MSSM, and is yet to be
demonstrated to occur in string theory); it can account for the suppression of
FCNC interactions in a natural way; in the minimal model (MSGM) it depends on
only four additional
parameters and one sign to describe all the masses and interactions of the 32
new SUSY particles.  (This may be compared with 110 new parameters
that can occur in the MSSM.)

The supergravity interactions of the MSGM produce four supersymmetry soft
breaking terms at $M_{G}$ scaled by $m_{o}$ (universal spin zero mass),
$m_{1/2}$ (universal gaugino mass), and $A_{o}$ and $B_{o}$ (cubic and
quadratic soft breaking constants).  One of the remarkable features of this
theory is that this spontaneous breaking of supersymmetry at $M_{G}$ generates,
by radiative renormalization group (RG) corrections, the breaking of
SU(2)$\times$U(1) at the electroweak scale [4]:  supersymmetry breaking
produces SU(2)$\times$U(1) breaking.  We will see below that radiative breaking
is a key element in the analysis of dark matter event rates, and failure to
include it loses much of the predictive power of the theory.

While the MSGM possesses only four additional parameters to describe SUSY
phenomena, and this is far fewer than the 20 to 30 of the 110 possible new
parameters that is commonly assumed in the MSSM, it still possesses a large
parameter space.  (Ideally, one would like to have four experiments to
determine
the four parameters, making all further predictions of the theory unique.)
Recently, however, there have been two new pieces of data, the CLEO measurement
of the $b\rightarrow s + \gamma$ branching ratio [5] and the CDF and DO
measurements of the top quark mass [6].  We will see below that while large
error
flags still remain in these data, they greatly reduce the allowed parameter
space.

There has been considerable activity in the recent past to calculate expected
event rates for dark matter detectors [7-13].  However, Refs. [7-9] do not
impose
radiative breaking and thus can get abnormally high event rates (often by
chosing the PC odd Higgs to be too light).  The major part of the analysis of
Ref. [10] is also done in this framework, and when radiative breaking is
introduced it is only for the special parameter choice $B_{o}=A_{o}-m_{o}$, and
the entire parameter space is not scanned. Thus Ref. [10] predicts
event rates that are too low.  The analyses [7-11] also impose relic
density constraints which leave out the full thermal averaging over the Z
and h (light Higgs) s-channel poles.  It is known that such omissions can
generate serious errors in density calculations [14,15], and
we will see below that it is important to treat s-channel poles correctly for
about half the parameter space.  Finally Refs. [7,8] do not include the heavy
Higgs, H, in the event rate calculation, the importance of which was first
pointed out in Ref. [16].

The plan of this paper is the following.  In Sec. 2 we briefly review the ideas
of radiative electroweak breaking and discuss the origin of the scaling
relations between the masses of the light neutralinos and charginos and the
gluino.  In Sec. 3 we exhibit an approximate analytic formula for the
gaugino and higgsino content of the $\tilde{Z}_{1}$ in the scaling regime.  In
Sec. 4 we discuss the relic density calculation, exhibit the importance of
correct treatment of the s-channel resonances, and also discuss the event rate
calculations.  We also show that uncertainties in the nucleon spin content do
not
have any significant effect on event rates for all targets except for the
lightest ones such as $^3He$ and $CaF_2$.  These results are in contrast with a
recent analysis [17] where the coherent part of the scattering was ignored and
hence claimed a large effect.  Sec. 5 is concerned with constraints on event
rates and SUSY parameter space due to the $b\rightarrow s + \gamma$ decay and
the top quark mass.  We show there that there are sizable regions of the
parameter space with R$>$ 0.01 including this constraint, and thus our results
differ from those of Ref.[18] which concludes that with the $b\rightarrow s +
\gamma$ constraint, the event rate R $< 0.01$.  We discuss in Sec. 6 the effect
of varying the endpoints of Eq. (3).  Sec. 7 gives the conclusions.  The
MSGM predicts that the  $\tilde{Z}_{1}$ is the lightest supersymmetric particle
over almost all the parameter space.  The analytic analysis of this is given in
the Appendix.
{}~\\
{}~\\
{\bf 2.~~Radiative Electroweak Breaking}\\
{}~\\
\indent
At the GUT scale the MSGM can be described by the superpotential
\begin{equation}
W=\mu_{o} H_{1}H_{2}+W_{Y} + {1\over{M_{G}}}W^{(4)}
\end{equation}
\noindent
where $W_{Y}$ is the cubic Yukawa couplings and $W^{(4)}$ contains any quartic
non-renormalizable couplings (which possibly lead to proton decay).  The
spontaneous breaking of supersymmetry leads to the soft supersymmetry breaking
effective potential $V_{SB}$ and gaugino mass term $L^{\lambda}_{mass}$,
\begin{equation}
V_{SB}=m_{o}^{2}\Sigma_{a}z_{a}^{+}z_{a}+(A_{o}W_{Y}+B_{o}\mu_{o}H_{1}H_{2}+h.c)
\end{equation}
\begin{equation}
L^{\lambda}_{mass} =-m_{1/2}\bar{\lambda}^{\alpha}\lambda^{\alpha}
\end{equation}
\noindent
where $\{z_{a}\}$ are the scalar fields and $\lambda^{\alpha}$ the gaugino
fields.  Equations (5-7) arise after the supergravity interactions cause the
spontaneous breaking of supersymmetry in the hidden sector, the GUT
interactions cause the breaking of the GUT group G to
SU(3)$\times$SU(2)$\times$U(1), all superheavy and hidden sector fields are
integrated out, and non-renoralizable terms scaled by (1/M$_{Pl}$) are
neglected.  The universality of the soft breaking parameters, $m_{o}, A_{o},
B_{o}$ and $m_{1/2}$, is a consequence of the universality of gravitational
couplings, plus the additional assumption that the hidden sector fields in the
Kahler potential also couple universally to the physical fields.  (This
universality also quarantees suppression of FCNC interactions and thus is
phenomenologically desirable).  The above results are generally insensitive to
the
nature of the GUT group provided the representations used to break G are not
too
large (so that GUT threshold corrections are not too big).

The effective potential may be reduced to the electroweak scale by using the
renormaliztion group (RG) equations.  Minimizing the Higgs potential with
respect to $<H_{1,2}>$ yields [4]
\begin{equation}
\mu^{2}={{\mu_{1}^{2}-\mu_{2}^{2}tan^{2}\beta}\over{tan^{2}\beta -1}}-{1\over
2}M_{Z}^{2};~~ sin2\beta = {{-2B\mu}\over{2\mu^{2}+\mu_{1}^{2}+\mu_{2}^{2}}}
\end{equation}
\noindent
Here  $\mu_{i}^{2}=m_{H_{i}}^{2}+\Sigma_{i}$ where $m_{H_{i}}^{2}$ is the
running $H_{i}$ mass at scale Q $\approx M_{Z}$ and $\Sigma_{i}$ are loop
corrections.  The $m_{H_{i}}^{2}$ are given by
\begin{equation}
m_{H_{1}}^{2}=m_{o}^{2}+m_{1/2}^{2} g(t)
\end{equation}
\begin{equation}
m_{H_{2}}^{2}=m_{1/2}^{2}e(t)+A_{o}m_{1/2}f(t)+m_{o}^{2}h(t)-k(t)A_{o}^{2}
\end{equation}
\noindent
where the form factors e,f,g,h,k are defined in Iba\~{n}ez et al [4],
the gluino mass is $m_{\tilde{g}}=(\alpha_{3}/\alpha_{G})m_{1/2}$ (with
$\alpha_{G}$ the GUT coupling constant), and t=ln($M_{G}^{2}/Q^{2}$).  One may
show that solutions exist to Eqs. (8), i.e. that SU(2)$\times$U(1) is
spontaneously broken, if and only if at least one of the supersymmetry soft
breaking interactions are non-zero.  Thus it is the supergravity interactions
at
$M_{G}$ that give rise to the breaking of
SU(2)$\times$U(1) at the electroweak scale.\\
\indent
To obtain a qualitative picture of the implications of electroweak breaking,
one notes from Eq. (8) that for $tan^{2}\beta >>1$ (i.e. $tan\beta
\stackrel{>}{\sim} 2-3$) that
\begin{equation}
\mu^{2}\simeq
-m_{H_{2}}^{2}-{1\over{2}}M_{Z}^{2}-\Sigma_{2}
\end{equation}
\noindent
Thus for electroweak breaking to occur, it is necessary for $m_{H_{2}}^{2}$ to
turn negative.  ($\Sigma_{2}$ is generally a small correction).  The
measured value of the top quark mass is $m_{t}=(176\pm 8\pm 10)$ GeV from CDF
[6] and $m_{t}=(199_{-21}^{+19}\pm 22)$ GeV from DO [6] while indirect
determinations from LEP yield $m_t=164_{-10}^{+9}{+6}_{-4}$ GeV [19].
These imply that
the top is relatively close to its Landau pole.  In this domain, for Q at the
electroweak scale, h(t), e(t) and f(t) are negative and g(t) and k(t) are
positive. Further, both $\mid h\mid$ and $\mid e\mid$ are O(1).  (For a
detailed
discussion of the effects of the Landau pole see Ref. [20].)  Thus
$m_{H_{2}}^{2}$ does indeed turn negative allowing SU(2)$\times$U(1) breaking
to
occur at the electroweak scale.

In the following we will restrict $m_{o}$ and $m_{\tilde{g}}$ to be less than 1
TeV to prevent an unreasonable amount of fine tuning.  [Actually we will see
that Eq. (4) actually implies $m_{\tilde{g}}\stackrel{<}{\sim}$(650-700)
GeV.].  Eq. (11) then implies
\begin{equation}
\mu^{2}>>M_{Z}^{2}
\end{equation}
\noindent
(which is satisfied if $\mu \stackrel{>}{\sim}(2-3) M_{Z}$) for almost the
entire
parameter space .  Note also that the above discussion implies that $\mu^{2}$
is an
increasing function of $m_{\tilde{g}}^{2}$ and
$m_{o}^{2}$.

Eq. (12) is essentially the remnants of the gauge hierarchy problem in
supersymmetry, i.e. from Eqs. (9-11), $\mu$ is scaled by $m_{o}$ and
$m_{\tilde{g}}$ and we allow the latter to go as high as 1 TeV.  It has been
previously
shown [21] that Eq. (12) leads to a set of scaling relations between the light
neutralinos, chargino and gluino,
\begin{equation}
2m_{\tilde{Z}_{1}}\cong
m_{\tilde{Z}_{2}}\cong m_{\tilde{W}_{1}}\simeq ({1\over
3}-{1\over4})m_{\tilde{g
}}
\end{equation}
\noindent
as well as the additional relations
\begin{equation}
m_{\tilde{Z}_{3,4}}\cong m_{\tilde{W}_{2}}>>m_{\tilde{Z}_{1}};~~ m_{H}\cong
m_{{H}^{\pm}}\cong m_{A}>>m_{h}.
\end{equation}
\noindent
Also one finds
\begin{equation}
63GeV\leq {m_h}\stackrel{<}{\sim} 120 GeV
\end{equation}

\noindent
where the lower bound on $m_h$ is the current LEP limit.
Eqs. (12-14) are thus a direct consequence of radiative electroweak breaking,
and we will see below that they play a dominant role in determining relic
densities and dark matter detection rates.\\
{}~\\
\noindent
{\bf 3.~~Composition of $\tilde{Z}_{1}$}\\
{}~\\
\indent
The $\tilde{Z}_{1}$ is generally a mixture of gauginos $\tilde{W}_{3},
\tilde{B}$ and Higgsinos $\tilde{H}_{1}, \tilde{H}_{2}$.  We write
\begin{equation}
\tilde{Z}_{1}=n_{1}\tilde{W}_{3}+n_{2}\tilde{B}+n_{3}\hat{H}_{1}+n_{4}
\tilde{H}_{2}
\end{equation}
\noindent
The expansion coefficients $n_{i}$ are determined by diagonalizing the
neutralino mass matrix.  In the ($\tilde{W}_{3}, \tilde{B}, \tilde{H}_{1},
\tilde{H}_{2}$) basis this reads [3]

\begin{equation}
M_{\tilde{Z}}=\pmatrix{\offinterlineskip
{\tilde{m}_{2}}&o&\vrule\strut&a&b\cr
o&{\tilde{m}_{1}}&\vrule\strut&c&d\cr
\noalign{\hrule}
a&c&\vrule\strut&o&{-\mu}\cr
b&d&\vrule\strut&{-\mu}&o\cr}
\end{equation}

\noindent
where $\tilde{m}_{i}=(\alpha_{i}/\alpha_{3})m_{\tilde{g}},~ a=M_{Z}
cos\theta_{W}cos\beta,~ b=-M_{Z}cos\theta_{W}sin\beta,~
c=-M_{Z}sin\theta_{W}cos\beta$ and $d=M_{Z}sin\theta_{W}sin\beta$.  The
$\tilde{Z}_{1}$ is the lowest mass eigenvector of $M_{\tilde{Z}}$, which may be
determined in general numerically.  However, as discussed in Sec. 2, radiative
breaking implies $\mu^{2}>>M_{Z}^{2}$.  One may generate an approximate
analytic form by treating the two blocks proportional to $M_{Z}$
perturbatively.  To second order perturbation theory one finds

\begin{eqnarray}
n_{1} & \cong &
-{1\over2}{M_{Z}\over\mu}{1\over{(1-\tilde{m}_{1}^{2}/\mu^{2})}}{M_{Z}\over
{\tilde{m}_{2}}-\tilde{m}_{1}}
sin2\theta_{W}\left[sin2\beta +{\tilde{m}_{1}\over\mu}\right]\\
n_{2} & = &1-{1\over2}{M_{Z}^{2}\over\mu^{2}}{1\over{(1-\tilde{m}_{1}^{2
}/\mu^{2})^{2}}}
sin^{2}\theta_{W}\left[1+{\tilde{m}_{1}\over\mu} sin2\beta
+{{\tilde{m}_{1}^{2}}
\over
\mu^{2}}\right]\\
n_{3} & = &{M_{Z}\over\mu}{1\over{1-\tilde{m}_{1}^{2}/\mu^{2}}}
sin\theta_{W}sin\beta\left[1+{\tilde{m}_{1}\over\mu} ctn\beta\right]\\
n_{4} & = &-{M_{Z}\over\mu}{1\over{1-\tilde{m}_{1}^{2}/\mu^{2}}}
sin\theta_{W} cos\beta\left[1+{\tilde{m}_{1}\over\mu} tan\beta\right]
\end{eqnarray}
\noindent
Eqs. (18-21) differ from the numerical computer results by amounts $\delta
n_{i}\stackrel{<}{\sim} 0.03$ over almost the entire parameter space (and
are generally a good deal better).  Thus one may use them to understand the
nature of the solutions.  One sees first that the $\tilde{Z}_{1}$ is mostly
bino since $n_{2}$ deviates from unity by second order effects,
$O(M_{Z}^{2}/\mu^{2})$.  Usually, $n_{2} > 0.95$ and often larger.  However,
$n_{3}, n_{4}$ and $n_{1}$ are first order, $O(M_{Z}/\mu)$.  This allows the
Higgsino components of the $\tilde{Z}_{1}$ to become considerable e.g.
$n_{3}\approx 0.2$.  The coherrent part of the $\tilde{Z}_{1}$-nucleus
scattering in dark matter detectors, depend upon the interference between the
gaugino and Higgsino components of the $\tilde{Z}_{1}$ [7].  Thus such terms
can become quite large, and as will be discussed in Sec. 4, this means that
the coherrent scattering almost always dominates the incoherrent (spin
dependent) scattering.  Thus there is a large difference between a
$\tilde{Z}_{1 }$ whose bino amplitude is 0.95 and one which is 100\% bino, and
one
cannot approximate the former as being pure bino.  We also note that the above
results are a direct consequence of the radiative breaking conditions, and
would
not in general hold without them.\\
{}~\\
{}~\\
\noindent
{\bf 4.~~Relic Density and Event Rate Analysis}\\
{}~\\
The primordial ${\tilde {Z}_{1}}$, created at the time of the Big Bang, can
annihilate in the early universe.  The main diagrams for this are shown in Fig.
1 [22].  At high temperature, the $\tilde{Z}_{1}$ is in thermal equilibrium
with its decay products.  However, when the annihilation rate falls below the
expansion rate of the universe, freeze out occurs at temperature $T_f$.  The
$\tilde {Z}_{1}$ are then disconnected from the background and continue to
annihilate.  In the simplest approximation [23] (accurate to about $\pm$ 25 \%)
the current relic density is given by

\begin{equation}
\Omega_{\tilde{Z}_{1}}h^2\cong 2.5\times
10^{-11}\left ({T_{\tilde{Z}_{1}}\over T_\gamma}\right )^3 {\left
(T_\gamma\over 2.75\right )^3}
 {\left(N_f\right )^{1\over 2}\over J(x_f)}
\end{equation}

\noindent
where
\begin{equation}
J(x_f)={\int_o^{x_f}}dx<\sigma v>;~~x_f=kT_f/m_{\tilde{Z}_{1}}
\end{equation}
\noindent
Here x=kT/m$_{\tilde{Z}_{1}}$, N$_f$ is the effective number of degrees of
freedom at freeze out, $(T_{\tilde{Z}_{1}}/T_\gamma)^3$ is the reheating
factor, and $T_\gamma$ the current microwave radiation temperature (in
$^\circ$K).  In Eq. (23), $\sigma$ is the annihilation cross section, v is the
relative velocity and $<~>$ means thermal average.

Freeze out generally occurs when the ${\tilde {Z}_{1}}$ are non-relativisitic,
i.e. at $x_f= kT_f/m_{\tilde{Z}_{1}}\simeq{1\over 20}$.  Thus the thermal
average may be taken over a Boltzmann distribution:
\begin{equation}
<\sigma v>=\int_0^{\infty} dvv^2\sigma v~exp[-v^2/4x]/\int_0^{\infty} dvv^2
exp[-v^2/4x]
\end{equation}
\noindent
The non-relativistic nature of the annihilation process has lead, in the past,
to making a non-relativistic expansion of $\sigma v$ i.e. $\sigma v\cong
a+b(v^2/c^2)+\cdots$; after which the thermal average becomes trivial to take.
This is a good approximation for the t-channel pole diagrams of Fig. 1, and for
the s-channel diagrams when 2m$_{\tilde {Z}_{1}}$ is not in the vicinity of
the h or Z poles.  However, when 2m$_{\tilde{Z}_{1}}$ is near the s-channel
poles, the non-relativistic approximation fails [14] and can produce errors as
large a factor of 100 [15] due to the fact that $\sigma$v is a rapidly varying
function in this region.  Further, the thermal averaging smears the region
where this
effect can occur (characteristically over a region $\approx$ 10 GeV in
m$_{\tilde{Z}_{1}}$ or $\approx$ 50 GeV in $m_{\tilde{g}}$).  For this
situation
one can first perform the integral of Eq. (24) analytically, and then calculate
Eq. (23) numerically.

In the following, we will investigate the parameter space defined by the
following bounds:

\begin{equation}
100 GeV\leq m_o\leq 1 TeV;~~ 150 GeV\leq m_{\tilde{g}}\leq 1 TeV
\end{equation}
\begin{equation}-6\leq A_t/m_o\leq 6;~~ 2\leq tan\beta\leq 20
\end{equation}
\noindent
The lower bound on m$_{\tilde{g}}$ is the current Tevatron bound  and the
upper bounds on m$_o$, m$_{\tilde{g}}$ are to prevent excessive fine tuning of
parameters.  (Similarly, tan$\beta$ $\stackrel{>}{\sim}$ 30 represents a fine
tuning of the Higgs VEV ratio.)  The range on A$_t$ covers the allowed
parameter
space (when all other experimental constraints are imposed).  One may estimate
the region of this parameter space where the above discussion of s-channel
poles
is important.  The scaling relations, Eqs. (13,14), allow us to choose
m$_{\tilde{g}}$ as the independent variable.  One sees from these relations,
that, varying over the field parameter space, one is generally near an
s-channel resonance (h or Z pole) when m$_{\tilde{g}}\stackrel{<}{\sim}$ 450
GeV, and
hence for this region one must treat the s-channel terms accurately.  This is
borne out by detailed numerical calculations which show that significant errors
occur in the calculation of $\Omega_{\tilde{Z}_{1}}h^2$ with
$m_{\tilde{g}}\leq$
450 GeV when the non-relativistic approximation to $<\sigma v>$is made, while
the approximation is generally good for m$_{\tilde{g}}>$ 450 GeV.  Since the
constraint on detection rate, R$\geq$ 0.01 events/kg da, requires
m$_{\tilde{g}}
{\stackrel{<}{\sim}}$ (650-700) GeV, we see that a correct treatment of the
s-channel poles is important for a large fraction of the total parameter space.

Detection of dark matter impinging on the Earth depends
upon the ${\tilde{Z}_{1}}$-quark scattering cross section for the quarks in the
nuclei of the detector.  The basic diagrams are shown in Fig. 2.  This process
has been studied by a number of authors [24], and can be represented by the
following effective Lagrangian:

\begin{equation}
L_{eff}=({\bar{\chi}_1}{\gamma^{\mu}}\gamma^5\chi_1)[\bar{q}\gamma_{\mu}(A_qP_L+B_qP_R)q]+
(\bar{\chi}_1 \chi_1)(\bar{q}C_qm_qq)
\end{equation}
\noindent
Here $\chi_1$ is the ${\tilde{Z}_{1}}$ field, q is the quark field and
P$_{R,L}$
= (1$\pm\gamma^5$)/2.  The coefficients A$_q$ and B$_q$ arise from the Z
t-channel pole and the ${\tilde{q}}$ s-channel pole, while C$_q$ comes
from the
$h^\circ$ and $H^{\circ}$ t-channel poles and the ${\tilde{q}}$ s-channel
pole.
(We follow the notation of Ellis and Flores [7] where explicit formulae for
A$_q$, B$_q$, C$_q$ are given for the ${\tilde{q}}$, h and Z pole diagrams.)

The first term of Eq. (27) gives rise to spin dependent (incoherrent)
scattering while the second term gives rise to spin independent (coherrent)
scattering.  Summing over all the quarks in the nucleus, the latter term then
produces an additional factor of the nuclear mass, M$_N$.  The event rate for
a detector then takes the form [24]

\begin{equation}
R=\left[ R_{SI}+R_{SD}\right ] \left [{\rho_{\tilde{Z}_{1}}\over 0.3GeV
cm^{-3}}\right ] \left [{v_{\tilde{Z}_{1}}\over 320 km/s}\right ]{events\over
kg~ da}
\end{equation}
\noindent
where ${\rho_{\tilde{Z}_{1}}}$ is the local mass density of
${\tilde{Z}_{1}}, v_{\tilde{Z}_{1}}$ the incident ${\tilde{Z}_{1}}$ velocity
and
the spin independent (SI) and spin dependent (SD) rates have the form [25]

\begin{equation}
R_{SI}={{16m_{\tilde{Z}_{1}}M_N^3}M_Z^4\over\left
[{M_N+m_{\tilde{Z}_{1}}}\right]^{2}}\,{\mid A_{SI}\mid^2}
\end{equation}

\begin{equation}
R_{SD} = {{16 m_{\tilde{Z}_{1}}}M_N\over \left
[{M_N+m_{\tilde{Z}_{1}}}\right ]^2}\,\lambda^2J(J+1)\mid A_{SD}\mid^2
\end{equation}
\noindent
where J is the nuclear spin and $\lambda$ is defined by $<N\mid\sum
{\stackrel{\rightarrow}{S}_i}\mid N>=\lambda<N\mid{{\stackrel{\rightarrow}
{J}}}\mid N>$.  We note that for large M$_N$, R$_{SI}\sim M_N$ while
$R_{SD}\sim 1/M_N$, showing that the heavy nuclei are best for detectors
sensitive to spin independent scattering.

Since the squarks are generally heavy over most of the parameter space,
A$_{SI}$
is dominated by its Higgs contributions which has the general form [26]

\begin{equation}
A^{Higgs}_{SI}\sim {g_2^2\over 4M_W}   \left[ {F_h\over m_h^2}
 \left\{
{{cos\alpha_H\over sin\beta} \atop {-sin\alpha_H\over cos\beta} }
\right\}
 + {F_H\over m_H^2}
\left\{
 {{sin\alpha_H\over sin\beta}\atop{cos\alpha_H\over cos\beta}}
\right\}
  \right ]{u-quark\atop d-quark}
\end{equation}
\noindent
where $F_h =(n_1-n_2 tan\theta_W)$$(n_4 cos\alpha_H$ +$n_3 sin\alpha_H)$ and
$F_H = (n_1-n_2tan\theta_W)$ $(n_4 sin\alpha_H-n_3 cos\alpha_H)$ and
$\alpha_H$ is the rotation angle needed to diagonalize the h-H$^{\circ}$ mass
matrix.  Thus the SI scattering arises from interference between the gaugino
and Higgsino parts of the ${\tilde{Z}_{1}}$ i.e. from Eqs. (18-21) from the
$n_2\times n_3$ terms of F$_{h,H}$ for most of the parameter space.  In
general (including the loop corrections to the Higgs mass matrix [27]) one
finds that $tan \alpha_H\approx$ 0.1.  Hence, for most of the parameter space,
the h contribution to d-quark scattering is suppressed by a factor of
tan$^2\alpha_H$ relative to the H, and this can overcome the fact that
$m_H~^2/m_h~^2>>1$.  In fact we find that the H contribution varies from 1/10
to 10 times the h contribution as one varies over the full parameter space,
and it is essential to keep both neutral CP even Higgs bosons in the analysis
[16].

In contrast to the above, the contributions to A$_{SD}$ from the Z pole
depends on $n_3^2-n_4^2$ which is small by Eqs. (18-21).  Thus for heavy
nuclei detectors one always has

\begin{equation}
R_{SI}>> R_{SD}
\end{equation}
\noindent
and even for the light CaF$_2$ detector which has a large $\lambda^2J(J+1)$
value, the spin independent scattering dominates over most of the parameter
space.

R$_{SD}$ depends on the spin content of the nucleons defined by

\begin{equation}
<n\mid{\bar{q}}\gamma^{\mu}\gamma^5q\mid n>=2s^{\mu}_{(n)}\Delta q
\end{equation}

\noindent
where $s^{\mu}_{(n)}$ is the spin 4-vector of nucleon n, and $\Delta$q measures
the part of the nucleon spin carried by quark q.  It has been suggested [17]
that
the differences between older data [28] and more recent data [29]
determinations of $\Delta$q (particularly $\Delta$s) could produce
uncertainties
in the value of R leading to errors as large as a factor of 30.  This could
indeed be the case if the ${\tilde{Z}_{1}}$ was pure gaugino (i.e. $n_2=1,
n_3=n_4=n_1=0$) for then R$_{SI}$ would vanish.  However, as we saw above, a
significant interference between the gaugino and Higgsino parts of the
${\tilde{Z}_{1}}$ exists (i.e. the ${\tilde{Z}_{1}}$ amplitude has a gaugino
amplitude of only about 0.95) leading instead to Eq. (32) for heavy
nuclei, minimizing the effect.  Fig. 3 shows the ratio between the predicted
value of the total R using the new and old data for $\Delta$q for a Ge
detector.  The difference is less than 10\% over the entire parameter space.
While larger errors can exist for a CaF$_2$ detector, where R$_{SD}$ is large,
even here the difference is less than $\pm$ 30\% over 92\% of the parameter
space.

A more serious uncertainty exists in R$_{SI}$ due to a lack of knowledge of
the s-quark content of a nucleon defined by $<n\mid m_s{\bar{s}}s\mid n>\equiv
f_sM_n$.  Estimates of f$_s$  [30] have about a 50\% uncertainty leading to a
$\pm$ (30-50) \% uncertainty in R$_{SI}$.  This will not, however, change the
qualitative nature of the results given below.

One may now understand analytically the dependence of the event rate on the
different SUSY parameters.  From Eq.(11), we see that $\mu^2$ is an
increasing function of -m$_{H_{2}}^2$ and hence by Eq. (10) an increasing
function of m$_{\tilde{g}}$ and m$_o$.  As shown in the discussion following
Eq. (31), the major contribution to A$_{SI}$ (which dominates the total event
rate R) is proportional to ($n_2n_3$) [or ($n_2n_4)$] which by Eqs. (18-21)
are the leading [O(M$_Z/\mu$] terms.  These decrease with increasing $\mu$.
Thus one expects R to be a decreasing function of m$_{\tilde{g}}$ and m$_o$.
This indeed was what was seen in the detailed computer calculations of Ref.
[12].  Further, for radiative breaking, which implies tan $\beta > 1$, Eq.
(31) also shows that the d-quark amplitude is an increasing function
of $tan\beta$.  Again the computer calculations of  Ref. [12] show this rapid
rise of R with tan $\beta$.

The calculation of the detector event rates now proceeds as follows.  One
calculates $\Omega_{\tilde{Z}_{1}}h^2$ and selects that part of the parameter
space of Eqs. (25,26) that satisfies the constraint of Eq. (3)  as well as the
LEP and Tevatron bounds on SUSY masses.  One then calculates the event rate R
for
this allowed part of the parameter space.  Figs. 4 and 5 show the maximum and
minimum event rates for $\mu < 0$ and $\mu >0$ as a function of A$_t/m_o$ for
Ge and CaF$_2$ detectors as one lets the remaining parameters vary over the
allowed parameter space.  (The relevant parts of these graphs are for the
regions where R$_{max} \geq 0.01.$)  As can be seen from the previous
discussion, the largest event rates occur for the largest allowed values of tan
$\beta$ and for the smallest values of m$_{\tilde{g}}$.  (The sharp peaks and
dips in the R$_{max}$ curves arise from the fact that a small value of
m$_{\tilde{g}}$ implies by Eq. (13) a small m$_{\tilde{W}_{1}}$.  If this
parameter point also satisfies the LEP cut that m$_{\tilde{W}_{1}} >$ 45 GeV it
is allowed and gives rise to a large event rate.  Otherwise it is excluded.)
The Ge detector is generally considerably more sensitive then the CaF$_2$
detector since its nuclear mass is considerably larger increasing the R$_{SI}$
contribution [as seen from Eqs. (29, 30)].

Figs. 4 and 5 also show that the parameter space is bounded in A$_t$ with most
of the allowed aregion occurring for$ A_t > 0$.  This phenomenon is due mainly
to the fact that the t quark mass is large and hence close to its Landau pole.
As one approaches the Landau pole, the light stop mass, $m_{\tilde{t}_{1}}$
obeys [20]

\begin{equation}
m_{\tilde{t}_{1}}^2=-{1\over 3}{A_R^2\over D_o}+m_{\tilde{t}_{1}}^2(NP)
\end{equation}
\noindent
where
\begin{equation}
A_R\cong A_t-0.613 m_{ \tilde{g}};~~ D_o\cong 0.164
\lbrack(m_t^f)^2-(m_t)^2\rbrack/M_W^2
\end{equation}
\noindent
$m_t^f\cong 197 sin \beta$ is the fixed point mass, and $m_{\tilde{t}_{1}}^2$
(NP) is a relatively smooth non-pole contribution.  For A$_R^2$ sufficiently
large, the ${\tilde{t}_{1}}$ becomes tachyonic, eliminating such parameter
points from the parameter space.  One expects then a lower bound on A$_t$ for
A$_t<$ 0, and similarly an upper bound for A$_t>$ 0 (as we are requiring
m$_{\tilde{g}} <1$ TeV), with the domain of positive A$_t$ being larger than
for
the negative  A$_t$ since cancelation in A$_R$ between the A$_t$ and
m$_{\tilde{g}}$ term can occur in the former case.  One finds, in fact,
including
correctly the non-pole part in Eq. (34), that for m$_t\simeq$ 170 GeV

\begin{equation}
-1.5~{\stackrel{<}{\sim}}~ A_t/m_o~{\stackrel{<}{\sim}}~ 5.5
\end{equation}
\noindent
Thus the high mass of the t quark eliminates a large amount of the SUSY
paramter space.
{}~\\
{}~\\
{\bf 5.~~Constraints From $b\rightarrow s+\gamma$ Decay}
{}~\\

The $b\rightarrow s+\gamma$ decay is a sensitive test for new physics, since it
is a flavor changing neutral current (FCNC) processs.  Thus the Standard Model
and new physics loops enter at the same order.  This is explicitly exhibited in
Fig. 6 where the basic diagrams for the decay at scale $\mu\approx M_W$ are
given
for SUSY models.  The W-t loop is common with the Standard Model, while the
other loops are the additional supersymmetric contributions.  The measured CLEO
branching ratio for $B\rightarrow X_s+\gamma$ is [5]

\begin{equation}
BR(B\rightarrow X_s\gamma))=(2.32\pm 0.5\pm 0.29 \pm 0.32)\times 10^{-4}
\end{equation}

\noindent
or combining all errors in quadrature one has BR(B$\rightarrow
X_s\gamma)\cong(2.32\pm 0.66)\times 10^{-4}$.  This result represents an
additional limitation on the allowed SUSY parameter space, and we discuss in
this section the effect the CLEO data has on dark matter detection event rates.

In the spectator approximation, the B meson decay can be related to the b quark
decay.  It is convenient, to define the quantity R as
\begin{equation}
{BR(B\rightarrow X_s\gamma)\over BR(B\rightarrow X_ce\bar{\nu}_e)}
\cong{\Gamma(b\rightarrow s+\gamma)\over\Gamma(b\rightarrow
c+e+{\bar{\nu}}_e)}\equiv R
\end{equation}
\noindent
(A CKM and $(m_b)^5$ factor, which have large errors, cancels out in R.).
Here BR(B$\rightarrow X_c e\bar{\nu}_ e$) = (10.7$\pm$ 0.5)\%.
The diagrams of Fig. 6 can be described by an effective Hamiltonian [31]

\begin{equation}
H_{eff}=V_{tb}V_{ts}^*{G_F\over\sqrt 2} C_7(M_W)Q_7
\end{equation}

\noindent
where $Q_7=(e/24\pi^2)m_b\bar{s}_L\sigma^{\mu\nu}b_RF_{\mu\nu}$.  Here
$F_{\mu\nu}$ is the electromagnetic field strength and $m_b$ is the b-quark
mass.  One must use the renormalization group equations to go from scale
$\mu = M_W$ to $\mu\simeq m_b$ where the decay occurs. To leading order (LO) in
QCD corrections, R is then calculated to be [31]
\begin{equation}
R=\mid {V_{tb}V_{ts}^*\over V_{cb}}\mid^2{6\alpha\over \pi I(z)}\mid
C_7^{eff}(m_b)\mid^2
\end{equation}
\noindent
where $I(z)=1 - 8 z^2+ 8z^6-z^8-24 z^4 \ell n z$ is a phase space factor
(z=m$_c/m_b$) and

\begin{equation}
C^{eff}_7(m_b)=\eta^{16\over 23} C_7(M_W)+{8\over 3}
(\eta^{14\over 23}-\eta^{16\over 23}) C_8(M_W)+C_2(M_W)
\end{equation}

\noindent
Here $Q_8=(g_3/16\pi^2) m_b{\bar{s}}_R\sigma^{\mu\nu}T^Ab_LG_{\mu\nu}^A$ where
T$^A$ and G$_{\mu\nu}^A$ are the gluon generators and field strengths and $C_2$
represents operator mixing with the 4-quark operators.

The QCD corrections are large for this process and the next to leading order
correction (NLO) are needed to obtain an accurate theoretical prediction.  At
present, however, not all the NLO terms have been calculated.  Thus the
theoretical analysis has an estimated error of about $\pm$ 30\% with a
Standard Model (SM) prediction of  $BR[B\rightarrow X_s\gamma]\cong (2.9\pm
0.8)\times 10^{-4}$ for m$_t$ = 174 GeV [32].  The H$^-$-t SUSY diagram of Fig.
6
adds constructively to the SM amplitude while the ${\tilde{W}} -{\tilde{t}}$
loop
may enter constructively or destructively with the SM model amplitude.  Since
the central value of the CLEO data of Eq. (37) already lies about 1 std. below
the central value of the SM results, the current data cannot tolerate a large
amount of constructive interference with SUSY amplitudes.  Thus, in spite of
the
large errors in the current data, the $b\rightarrow s+\gamma$ decay produces a
significant constraint on the SUSY parameter space.

As seen from Eqs. (14), radiative breaking generally implies $m_{H^{\pm}}$ is
large, suppressing its contribution to the $b\rightarrow s + \gamma$
decay.  The ${\tilde{t}}-{\tilde{W}}$ diagram can become large, however, when
m$_{\tilde{t}_{1}}$ and m$_{\tilde{W}_{1}}$ are small.  The
${\tilde{t}}~(mass)^2$ matrix reads
\begin{equation}
\left(
{ { {m_ {\tilde {t}_{L} }^{2}} \atop{m_{t}(A_{t}+\mu ctn\beta)} }
{ {m_{t}(A_{t}+\mu ctn\beta)} \atop {m_{\tilde{t}_{R}}^{2} } }}
\right)
\end{equation}

\noindent
where expressions for $m_{\tilde{t}_{L}}^2$, $m_{\tilde{t}_{R}}^2$ are given in
[20].  The light $\tilde{t}$ eigenvalue is
\begin{equation}
m_{\tilde{t}_{1}}^2 = {1\over 2}(m_{\tilde{t}_{L}}^2
+m_{\tilde{t}_{R}}^2)-\left [{1\over 4}
(m_{\tilde{t}_{L}}^2-m_{\tilde{t}_{R}}^2)^2 +m_t^2(A_t+\mu
ctn\beta)^2\right ]^{1\over 2}
\end{equation}
\noindent
Thus $m_{\tilde{t}_{1}}^2$ can become small if A$_t$ and $\mu$ can have the
same
sign and this effect is enhanced due to the fact that m$_t$ is so large [33].
One expects therefore that when A$_t$ and $\mu$ have the same sign the
theoretical prediction for BR(b$\rightarrow s+\gamma$) will become large if
also
m$_{\tilde{W}_{1}}$ is small, while the SUSY effect on BR (b$\rightarrow s
+\gamma$) will be small for A$_t$ and $\mu$ having opposite sign (or
m$_{\tilde{W}_{1}}$ becoming large).  This can be seen explicitly to be the
case
from detailed computer calculations given in [34].

Eq. (36) shows that the major part of the parameter space has A$_t >0$, and so
the b$\rightarrow s+\gamma$ decay is expected to influence mostly the $\mu>0$
branch.  One may quantify this by requiring that the theoretical rate for
BR(b$\rightarrow s+\gamma$) be within the 95 \% CL bounds of the experimental
value.  One finds then that for $\mu>0$ one is restricted to the region
\begin{equation}
-1.0 <A_t/m_o <0.5;~~~ \mu>0
\end{equation}

\noindent
i.e. about 40\% of the parameter space of Eq. (36) is eliminated by the
b$\rightarrow s+ \gamma$ data!  In addition, sections of the parameter space
with small $m_{\tilde{g}}$ are eliminated for A$_t < 0$, $\mu < 0$ and for $0<
A_t\leq 0.5$, $\mu >0$.

The effects of the above restrictions are shown in Fig. 7 ($\mu <$ 0) and Fig.
8 ($\mu > 0$).  In Fig. 7 one sees that event rates are reduced in the small
domain of A$_t < 0$, (as the low m$_{\tilde{W}_{1}}$ part of the parameter
space is eliminated), but not significantly modified over the remainder of the
parameter space where A$_t > 0$, since here $\mu$ and A$_t$ have opposite
signs.  Fig. 8 shows that the only remaining part of the parameter space is the
narrow band of A$_t < 0.5$, and all of the parameter space with A$_t > 0.5$
(where $\mu$ and A$_t$ have the same sign) is eliminated.  Thus the major
effect of the  b$\rightarrow s+\gamma$ data is to eliminate regions of
parameter
space, rather than reduce expected event rates.  As can be seen from Figs. 7
and 8, there still remain regions of parameter space with large event rates.
(This is to be contrasted with the results stated in [18].)
{}~\\
{}~\\
{\bf 6.~~Varying The Bounds On $\Omega_{\tilde{Z}_{1}}h^2$}
{}~\\
\indent
The analysis considered above was done within the framework of the bounds
of Eq. (3) on $\Omega_{\tilde{Z}_{1}}h^2$.  We consider now the effect of
varying
these bounds.  We first note that the $\tilde{Z}_{1}$ annihilation cross
section
in the early universe arising from the diagrams of Fig. 1 is a decreasing
function of m$_{\tilde{Z}_{1}}$.  Alternately, from the scaling relations Eq.
(13), one may say then that $\Omega_{\tilde{Z}_{1}}h^2$ increases as
m$_{\tilde{g}}$ increases.  On the other hand, we saw in Sec. 4 that the event
rate R is a decreasing function of m$_{\tilde{g}}$.  We are interested in this
paper in the region $R\geq$0.01 events/kg da which will be accessible
experimentally in the forseeable future.  This bound on R then puts an upper
bound on m$_{\tilde{g}}$.  The largest allowed value of m$_{\tilde{g}}$ occurs
for the largest value of tan$\beta$ and smallest value of m$_o$, (which by Eqs.
(25,26) we are taking as tan$\beta$$\leq$ 20, m$_o\geq$ 100 GeV), since R
increases with tan$\beta$ and decreases with m$_o$.  One finds then for the
parameter space defined by Eqs. (25,26) that for $\mu <0$ one has[35]

\begin{equation}
m_{\tilde{g}}\stackrel{<}{\sim} 650 GeV;~~ for~ R\geq 0.01,~\mu<0
\end{equation}

\noindent
For $\mu>0$, m$_{\tilde{g}}$ can rise to (700-750) GeV.  However, this occurs
for
$A_t>0$, and as discussed in Sec. 5, this region of the parameter space
is eliminated by the b$\rightarrow s+\gamma$ decay data, and so Eq. (45)
represents the true upper bound on m$_{\tilde{g}}$ for dark matter that can be
detected with current designs of detectors.

A detailed inspection of the full parameter space shows that the limitation
R$<$
0.01 implies then that the early universe annihilation channels
${\tilde{Z}_{1}}{\tilde{Z}_{1}}\rightarrow$ hh, Zh are closed.  The channel
${\tilde{Z}_{1}}{\tilde{Z}_{1}}\rightarrow$WW,ZZ is almost always closed, and
the small regions in parameter space which allow this annihilation are very
close to threshold and hence highly suppressed.  Hence, one need not consider
the vector meson channels in the analysis of $\Omega_{\tilde{Z}_{1}}h^2$ given
in Sec. 4.

The behavior of $\Omega_{\tilde{Z}_{1}}h^2$ and R as a function of
m$_{\tilde{g}}$ also shows that the bound R$\geq$ 0.01 implies that
$m_{\tilde g}\leq 650$ GeV. This is exhibited in  Fig.~9a.
One sees,as shown in Fig.~9b, that $\Omega_{\tilde{Z}_{1}}h^2 <$0.3
 when R$\geq$ 0.01 at these maximum values of $m_{\tilde g}$.
Thus in this domain, the results obtained above are not sensitive to the
precise upper bound in Eq. (3).  If one were to lower the upper bound below
$\Omega_{\tilde{Z}_{1}}h^2$ = 0.3, then the upper bound on m$_{\tilde{g}}$ is
further reduced.  This is shown in Fig. 10, where it is seen that if
$\Omega_{\tilde{Z}_{1}}h^2<0.2$, then m$_{\tilde{g}} <$ 400 GeV, which would
make the gluino accessible to the proposed high luminosity upgrade of the
Tevatron [i.e. the TeV(33)].  We note that in both the inflationary scenario
with a cold-hot dark matter mixture or the scenario with cold dark matter and
a cosmological constant, low values of $\Omega_{\tilde{Z}_{1}}h^2$
are preferred.  This follows from the fact that in the former case one needs
a small value of h (i.e. h $\simeq$ 0.5 so that the age of universe be
consistent with the estimated age of globular clusters) and in the latter
case because $\Omega_{\tilde{Z}_{1}}$ is small (i.e.
$\Omega_{\tilde{Z}_{1}}\simeq 0.2-0.4$ since the majority of the matter of the
universe is in the cosmological constant).

We now turn to the question of sensitivity of results to the lower bound on
$\Omega_{\tilde{Z}_{1}}h^2$.  As discussed above,
low $\Omega_{\tilde{Z}_{1}}h^2$ arises when m$_{\tilde{g}}$ is small, which is
also the domain of parameter space where R can be large.  Also, as discussed in
Sec. 4, the peaks and dips of $R_{max}$ in Figs. 4 and 5 arise from whether or
not the LEP bound m$_{\tilde{W}_{1}} > 45$ GeV can be satisfied.  Figs. 11 and
12 exhibit the effect on the maximum event rates for Pb and CaF$_2$ detectors
when the bound $\Omega_{\tilde{Z}_{1}}h^2 > 0.10$ is raised to
$\Omega_{\tilde{Z}_{1}}h^2 > 0.15$.  One sees that the sharp peaks get clipped
off, but otherwise the results are qualitatively unchanged.  There still
remains, however, a sizable amount parameter space where R exceeds 0.01.
{}~\\
{}~\\
{\bf 7.~~Conclusions}
{}~\\
\indent
We have examined in this paper the direct detection possibility of
${\tilde{Z}_{1}}$ cold dark matter within the framework of the minimal
supergravity model (MSGM) for the parameter space defined by Eqs. (25,26), and
for relic densities $\Omega_{\tilde{Z}_{1}}h^2$ in the range given by Eq.
(3). [Eq. (3) encompasses the range one would expect from inflationary
cosmology for either the cold-hot dark matter senario or the cold dark
matter plus cosmological constant possibility.]  One further limits the
parameter space so that experimental bounds on SUSY masses are obeyed.  Two
new pieces of data, the t quark mass and the b$\rightarrow s+\gamma$
branching ratio, have greatly constrained the SUSY parameter space.  Thus the
fact that m$_t$ is large (i.e. close to its quasi infra-red fixed point) limits
the domain of A$_t$ to be mostly positive, while the experimental
b$\rightarrow s+\gamma$ decay rate limits the parameter space to be mostly
where A$_t$ and $\mu$ have the opposite sign.  Thus the majority (though
not all) of the allowed parameter space is in the region where A$_t>$ 0 and
$\mu <$ 0.

Physical quantities in the MSGM depend on four SUSY parameters, m$_o$,
m$_{\tilde{g}}$, A$_t$, tan$\beta$, and the sign of $\mu$.  Thus a general
quantity has a complex behavior as one varies over the full parameter
space.  Radiative breaking, however, plays a central role in MSGM
predictions and allow one to understand analytically the qualitative behavior
of
the event rate for the detection of ${\tilde{Z}_{1}}$ dark matter.
Approximate analytic expressions were obtained in Sec. 3 for the content of
the ${\tilde{Z}_{1}}$ showing that the ${\tilde{Z}_{1}}$ was mostly bino,
but with a non-negligible amount of higgsino.  One can see from this that the
spin independent (coherrent) contribution generally dominates R (and hence
the most sensitve detectors are those with the heaviest nuclei) and that R
decreased with m$_o$ and m$_{\tilde{g}}$ and increases with tan$\beta$.

Current dark matter detectors hope to obtain a sensitivity of R$>$ 0.01
events/kg da in the forseeable future.  As seen here, such a sensitivity will
allow the exploration of a sizable amount of the SUSY parameter space,
though there will still remain large sections that fall below this bound.
The domain R $>$0.01 correpsonds to m$_{\tilde{g}}$
$\stackrel{<}{\sim}$ 650 GeV and is insensitive to the choice of
upper bound on $\Omega_{\tilde{Z}_{1}}h^2$ of Eq. (3) provided
$\Omega_{\tilde{Z}_{1}}h^2\stackrel{<}{\sim} 0.3$, and only mildly sensitive to
the lower bound of Eq. (3).  Thus dark matter could only be expected to be seen
by current detector designs if the gluino is not too heavy.
{}~\\
{}~\\
\noindent
{\bf Acknowledgements}
{}~\\
\indent
This research was supported in part by NSF grant numbers PHY-9411543,
PHY-19306906  and PHY94-07194.
\newpage
\noindent
{\bf Appendix}
{}~\\
{}~\\
\indent
In N=1 supergravity there are three possible candidates for the LSP.  These are
the ${\tilde{Z}_{1}}$, the ${\tilde{\nu}}$ and the ${\tilde{e}_{R}}$.  We
discuss here the region of parameter space where the ${\tilde{Z}_{1}}$ is the
LSP.  The basic requirement for this is then that

\begin{flushright}
$m_ {\tilde{e}_{R}} >m_{\tilde{Z}_{1}};~~m_{\tilde{\nu}}>m_{\tilde{Z}_{1}}
{}~~~~~~~~~~~~~~~~~~~~~~~~~~~~~~~~~~{(A.1)}$
\end{flushright}
\noindent
In addition one has the LEP constraint that
\begin{flushright}
$m_{\tilde{Z}_{1}}\stackrel{>}{\sim}20
GeV~~~~~~~~~~~~~~~~~~~~~~~~~~~~~~~~~~~~~~{(A.2)}$
\end{flushright}
\noindent
The requirement that the ${\tilde{e}_{R}}$ be heavier than ${\tilde{Z}_{1}}$
is an experimental constraint as a charged LSP would have already been
discovered.  We will see that in fact this condition is obeyed for almost the
entire parameter space.  The region where the ${\tilde{\nu}}$ is the LSP will
also be seen to be very small.

The ${\tilde{e}_{R}}$ and ${\tilde{\nu}}$ masses can be expressed in terms of
the basic MSGM parameters as [4]
\begin{flushright}
$m^2_{\tilde{e}_{R}} =m_o^2+ {\tilde\alpha}_{G}{6\over 5}f_1
m_{1\over 2}^2 - sin^2\theta_W
M_Z^2cos2\beta~~~~~~~~~~~~~~~~~{(A.3)}$
\end{flushright}
\begin{flushright}
$m_{\tilde{v}}^2=m_o^2+{\tilde{\alpha}_{G}}\left [{3\over 2}f_2+{3\over
10}f_1\right ] m^2_{1\over 2}+{1\over 2} M^2_Z
cos2\beta~~~~~~~~~~~~{(A.4)}$
\end{flushright}
\noindent
where the form factors $f_i(t)$ are defined as
\begin{flushright}
$f_i(t)=t~(2+\beta_it)/(1+\beta_it)^2 ~~~~~~~~~~~~~{(A.5)}$
\end{flushright}
\noindent
where
$t=2 \ell n (M_G/Q)$ and $\beta_i={\tilde{\alpha}}_G({33\over 5},~ 1,~ -3)$
are the $U(1)XSU(2)XSU(3)$ $\beta$-functions with
${\tilde{\alpha}_{G}}=\alpha_G/4\pi$.  In the following we use
$\alpha_G={1\over 24}, M_G=2\times10^{16}GeV, Q= M_Z$.  One finds then
$f_1\cong 38.1, f_2\cong 99.0, f_3\cong 772.0$.  In order to
analytically illustrate the results, we will perform the remainder of
the calculation in the scaling limit of Eq. (13,14).  (A more accurate
numerical calculation give results close to the analytic ones.)  One
can then relate $m_{1\over 2}$ to the ${\tilde{Z}_{1}}$ mass by
\begin{flushright}
$m_{\tilde{Z}_{1}}\cong {\tilde{m}_{1}}=(\alpha_1/\alpha_G)m_{1\over
2}~~~~~~~~~~~~~~~~~~~~~~~~~{(A.6)}$
\end{flushright}
\noindent
or $m_{1\over 2}\cong 2.45 m_{\tilde{Z}_{1}}$. Eqs. (A.3,A.4) then
become
\begin{flushright}
$m_{\tilde{e}_{R}}^2\cong m_o^2+0.912
m_{\tilde{Z}_{1}}^2-sin^2\theta_WM_Z^2cos2
\beta~~~~~~~~~~~~~~~~~~~~~~~~~{(A.7)}$
\end{flushright}
\begin{flushright}
$m_{\tilde{\nu}}^2\cong m_o^2+3.19 m_{\tilde{Z}_{1}}^2 +{1\over 2}
M_Z^2 cos 2\beta~~~~~~~~~~~~~~~~~~~~~~~{(A.8)}$
\end{flushright}
\noindent
The requirement that the $\tilde{e}_{R}$ not be
the LSP then becomes an upper bound on $m_{\tilde{Z}_{1}}$:
\begin{flushright}
$m_{\tilde{Z}_{1}}<11.4 \left[
m_o^2+sin^2\theta_W M_Z^2(-cos2\beta)
\right]~~~~~~~~~~~~~~~~~~~~~~~~~~{(A.9)}$
\end{flushright}
\noindent
Similarly, the condition that the ${\tilde{\nu}}$ is heavier than the
${\tilde{Z}_{1}}$ becomes a lower bound on $m_{\tilde{Z}_{1}}$
\begin{flushright}
$m_{\tilde{Z}_{1}}^2>0.456\left [-m_o^2+{1\over
2}M_Z^2(-cos2\beta)\right ]~~~~~~~~~~~~~~~~~~~~~~{(A.10)}$
\end{flushright}
\indent
Eqs. (A.9, A.10, A.2) are three inequalities constraining the
parameters $m_{\tilde{Z}_{1}}$, m$_o$  and tan$\beta$.  In addition we
have the fine tuning constraints of m$_o$, m${\tilde{g}}<$\\
1 TeV.

In the scaling limit then Eq. (A.9) implies for m$_{\tilde{g}}$ = 1 TeV
\begin{flushright}
$m_o>41.9,~for~ tan\beta=1;~~m_o\geq 0,~for~ tan\beta\geq
4.7~~~~~~~~~~~~~~~~~~~~~{(A.11)}$
\end{flushright}
\noindent
with corresponding smaller lower bounds on m$_o$ for smaller values of
m$_{\tilde{g}}$.  Thus for almost the entire parameter space the MSGM predicts
that the ${\tilde{e}_{R}}$ is not the LSP.

Turning to the ${\tilde{\nu}}$ constraint, we see that the r.h.s. of Eq. (A.10)
falls below the LEP bound Eq. (A.2) (and hence becomes irrelevant) when
\begin{flushright}
$m_o^2>{1\over 2} M_Z^2(-cos2\beta)-(20)^2/
0.456~~~~~~~~~~~~~~~~~~~~~{(A.12)}$
\end{flushright}
\noindent
and hence
\begin{flushright}
$m_o\geq 0 ~for~ tan\beta<1.24;~~ m_o>57.3 GeV ~for~
tan\beta>>1~~~~~~~~~~{(A.13)}$
\end{flushright}
\noindent
Alternately for all tan$\beta$, Eq.(A.10) is satisfied for all m$_o$ if
m$_{\tilde{Z}_{1}} >$ 43.6 GeV (or m$_{\tilde{g}}>$ 308 GeV).  Thus the
$\tilde{\nu}$ is predicted to be heavier than the ${\tilde{Z}_{1}}$ for all but
a very small portion of the parameter space.

\newpage
{\bf References}\\
\begin{enumerate}
\item
E. Gates, G. Gyuk and M. Turner, Phys. Rev. Lett. {\bf 74}, 3724
(1995).
\item
L.M. Krauss and M. Turner, CWRU-P6-95-FERMILAB-Pub-95/063-A;
astro-ph/9504003.
\item
A.H. Chamseddine, R. Arnowitt and P. Nath, Phys. Rev. Lett {\bf 29}.
970 (1982).  For reviews see ``Applied N=1 Supergravity" (World Scientific,
Singapore, 1984); H.P. Nilles, Phys. Rep. {\bf 110}, 1 (1984); R. Arnowitt and
P. Nath, Proc of VII J.A. Swieca Summer School (World Scientific, Singapore
1994).
\item
K. Inoue et al. Prog. Theor. Phys {\bf 68}, 927 (1982); L. Iba\~{n}ez and
G.G. Ross, Phys. Lett. {\bf B110}, 227 (1982); L. Avarez-Gaum\'{e},
J.Polchinski and M.B. Wise, Nucl. Phys. {\bf B221}, 495 (1983), J. Ellis, J.
Hagelin, D.V. Nanopoulos and K. Tamvakis, Phys. Lett. {\bf B125}, 2275 (1983);
L.E. Iban\~{n}ez and C. Lopez, , Nucl. Phys. {\bf B233}, 545 (1984); L.E.
Iba\~{n}ez, C. Lopez and C. Mu\~{n}os, Nucl. Phys. {\bf B256}, 218
(1985).
\item
M.S. Alam et al. (CLEO Collaboration), Phys. Rev. Lett. {\bf 74}, 2885 (1995).
\item
F. Abe et al. (CDF Collaboration) Phys. Rev. Lett. {\bf 74}, 2626 (1995); S.
Abachi et al. (DO Collaboration) Phys. Rev. Lett. {\bf 74}. 2632 (1995).
\item
J. Ellis and R. Flores, Phys. Lett. {\bf B263}, 259 (1991); {\bf B300}, 175
(1993); Nucl. Phys. {\bf B400}, 25 (1993).
\item
K. Greist, Phys. Rev. Lett. {\bf 61}, 666 (1988).
\item
A. Bottino et al, Astro. Part. Phys. {\bf 1}, 61 (1992); {\bf 2}, 77 (1994).
\item
M. Drees and M. Nojiri, Phys. Rev. {\bf D48}, 3483 (1993).
\item
G. Kane, C. Kolda, L. Roszkowski and J. Wells, Phys. Rev. {\bf D49}, 6173
(1994); E. Diehl, G. Kane, C. Kolda and J. Wells, UM-TH-94-38 (1994).
\item
R. Arnowitt and P. Nath, to be pub. Mod. Phys. Lett. {\bf A10}: 1257 (1995).
\item
P. Nath and R. Arnowitt, Phys. Lett. {\bf B336}, 395 (1994); Phys. Rev. Lett.
{\bf 74}, 4592 (1995).
\item
K. Greist and D. Seckel, Phys. Rev. {\bf D43}, 3191 (1991); P. Gondolo and G.
Gelmini, Nucl. Phys. {\bf B360}, 145 (1991).
\item
R. Arnowitt and P. Nath, Phys. Lett. {\bf B299}, 58 (1993); {\bf B303}, 403
(1993) (E); P. Nath and R. Arnowitt, Phys. Rev. Lett. {\bf 70}, 3696 (1993).
After completion of this paper there has appreared an analysis by H.Baer
and M.Brhlick, Florida State University Preprint FSU-HEP-950818, which
implements numerically the computation of the relic density without
using a power series expansion in velocity near an s-channel resonance.
The thermal averaging of this analysis is equivalent to ours.
\item
M. Kamionkowski, Phys. Rev. {\bf D44}, 3021 (1991).
\item
M. Kamionkowski, L. Krauss and M. Ressel, IASSN-HEP-94/14-CWRU-P3-94.
\item
F. Borzumati, M. Drees and M. Nojiri, Phys. Rev. {\bf D51}, 341 (1995).
\item
P. Langacker, talk at SUSY 95, Palaiseau, France, 1995.
\item
P. Nath, J. Wu and R. Arnowitt, NUB -TH-3116/95-CTP-TAMU-07/95.
\item
R. Arnowitt and P. Nath, Phys. Rev. Lett. {\bf 69}, 725 (1992); P. Nath and R.
Arnowitt, Phys. Lett. {\bf B289}, 368 (1992).
\item
As will be discussed in Sec. 6, for R$>$0.01 events/kg da, the annihilation
channels into vector bosons and Higgs bosons are effectively closed.
\item
B.W. Lee and S. Weinberg, Phys. Rev. Lett. {\bf 39}, 165 (1977); D.A. Dicus, E.
Kolb and V. Teplitz, Phys. Rev. Lett. {\bf 50}, 1419 (1983); H. Goldberg, Phys.
Rev. Lett. {\bf 50}, 1419 (1983); J. Ellis, J.S. Hagelin, D.V. Nanopoulos and
N. Srednicki, Nucl. Phys. {\bf B238}, 453 (1984).
\item
M.W. Goodman and E. Witten, Phys. Rev. {\bf D~31}, 3059 (1993); K. Greist,
Phys.
Rev. {\bf D38}, 2357 (1988); {\bf D39}, 3802 (1989)(E); R. Barbieri, M. Frigini
and G.F. Giudice, Nucl. Phys. {\bf B313}, 725 (1989); M. Strednicki and R.
Watkins, Phys. Lett. {\bf B225}, 140 (1989).
\item
A factor of 4 has been included in Eqs. (29,30) to account for the Majorana
Nature of the ${\tilde{Z}_{1}}$, in accord with Ref. [10].
\item
J.F. Gunion, H.E. Haber, G. Kane and S. Dawson, ``The Higgs Hunter's Guide"
(Addison-Wesley Publ Co., Redwood City, 1990).
\item
H. Haber and R. Hempfling, Phys. Rev. Lett. {\bf 60}, 1815 (1991); Phys. Rev.
{\bf D48}, 4280 (1993); J. Ellis, G. Ridolfi and F. Zwirner, Phys. Lett. {\bf
B262}, 477 (1991).
\item
J. Ashman etal (EMC Collaboration), Nucl. Phys. {\bf B328}, 1 (1989); A.
Manohar and R. Jaffe, Nucl. Phys. {\bf B337}, 509 (1990).
\item
D. Adams et al. (SMC Collaboration), Phys. Lett. {\bf B329}, 399 (1994); R.
Arnold et al. (E143 Collaboration) in ``Intersections of Particle and Nuclear
Physics", St. Petersburg, FL (1994).
\item
T.P. Cheng, Phys. Rev. {\bf D38}, 2869 (1989); J. Gasser, H. Leutwyler and M.E.
Saino, Phys. Lett. {\bf B253}, 252 (1991).
\item
S. Bertolini, F. Borzumati and A. Masiero, Phys. Rev. Lett. {\bf 59}, 180
(1987); B. Grinstein, R. Springer and M.B. Wise, Nucl. Phys. {\bf B339}, 269
(1990); S. Bertolini, F. Borzumati, A. Masiero and G. Ridolfi, Nucl. Phys. {\bf
B353}, 591 (1991); R. Barbieri and G. Giudice, Phys. Lett. {\bf B309}, 86
(1993); M. Misiak, Phys. Lett. {\bf B269}, 161 (1991); Nucl. Phys. {\bf B393},
23 (1993).
\item
A.J. Buras, M. Misiak, M. M\"{u}nz and S. Pokorski, Nucl. Phys. {\bf B424}, 374
(1994); M. Ciuchini, E. Franco, G. Martinelli, L. Reina and L. Silvestrini,
Phys. Lett. {\bf B316}, 127 (1993).
\item
m$_{\tilde{t}_{1}}~^2$ can also become small if m$_{\tilde{t}_{R}}~^2$ turns
negative.  This in fact is what happens when the t-quark is close to its Landau
pole, as discussed in Sec. 4.
\item
J. Wu, P. Nath and R. Arnowitt, Phys. Rev. {\bf D51}, 1371 (1994).
\item
We use here value of R for a Pb detector, since this detector is most sensitive
as it has the highest mass nuclei.
\newpage
\noindent
{\bf Figure Captions}\\
{}~\\
\noindent
Fig.~~1~~${\tilde{Z}_{1}}$ annihilation diagrams for annihilation in the
early universe.\\
{}~\\
\noindent
Fig.~~2~~${\tilde{Z}_{1}}$-quark scattering diagrams for a terrestial dark
matter detector.\\
{}~\\
\noindent
Fig.~~3~~R(New)/R(Old) vs $\mu$ for a Ge detector.  ``New" data is Ref. [29]
and
"Old" data Ref. [28].\\
{}~\\
\noindent
Fig.~~4~~Maximum and minimum event rates for Pb (solid) and CaF$_2$ (dashed)
detectors as a function of A$_t$ for $\mu < $0 and m$_t$ = 168 GeV. The other
parameters are varied over the range of Eqs. (25,26). (From Ref. [12].)\\
{}~\\
\noindent
Fig.~~5~~Same as Fig. 4 for $\mu >$ 0.\\
{}~\\
\noindent
Fig.~~6~~ Diagrams contribution to b$\rightarrow s+\gamma$ decay at W mass
scale.\\
{}~\\
\noindent
Fig.~~7~~Maximum and minimum event rates for Pb detector vs A$_t/m_o$ without
b$\rightarrow s+\gamma$ constraint (solid)  and with b$\rightarrow s +\gamma$
constraint (dot-dashed) at 95\% CL for $\mu<$ 0.  Other parameters are varied
over the range of Eqs. (25,26) with m$_t$=168 GeV. (The minimum event rates are
unaffected by the b$\rightarrow s+\gamma$ constraint.)\\
{}~\\
\noindent
Fig.~~8~~Same as Fig. 7 for $\mu>0$.  The parameter space with the
$b\rightarrow s + \gamma$ constraint terminates for A$_t/m_o >$ 0.5.\\
{}~\\
\noindent
Fig.~~9~~Maximum value of m$_{\tilde{g}}$ (Fig.~9a) and maximum value of
$\Omega_{\tilde{Z}_{1}}h^2$ at this $m_{\tilde g}$(Fig.~9b) as a
function of tan$\beta$ for $\mu <$
0 for domain R$ >$ 0.01 events/kg da for Pb detector as other parameters are
varied over range of Eqs. (25,26) with m$_t$ = 168 GeV.  The curves from bottom
to top are for A$_t/m_o$ = 0, 2.0, 4.0.\\
{}~\\
\noindent
Fig.~~10~~Maximum value of m$_{\tilde{g}}$ as a function of the upper bound on
$\Omega_{\tilde{Z}_{1}}h^2$ for $\mu<0$ for $tan\beta$ = 6, A$_t/m_o$ = 0.5.
(Results are insensitive to the values of $tan\beta$ and A$_t$.)\\
{}~\\
\noindent
Fig.~~11~~Maximum event rates for Pb detector as a function of $A_t/m_o$
for $\mu<$ 0 for $\Omega_{\tilde{Z}_{1}}h^2 >$ 0.10 (solid) and
$\Omega_{\tilde{Z}_{1}}h^2 >$ 0.15 (dot-dashed).  Other parameters are varied
over the range of Eq. (25,26) with m$_t$ =168 GeV.  (The b$\rightarrow
s+\gamma$
decay constraint is not here imposed.)\\
{}~\\
\noindent
Fig.~~12~~ Maximum event rates for CaF$_2$ detector as a function of A$_t/m_o$
for $\mu<$ 0 for $\Omega_{\tilde{Z}_{1}}h^2 >$ 0.10 (dashed) and
$\Omega_{\tilde{Z}_{1}}h^2 >$ 0.15 (solid).  Other parameters are varied over
the range of Eqs. (25,26) with m$_t$ = 168 GeV.  (The b$\rightarrow s+\gamma$
decay constraint is not here imposed.)\\
\end{enumerate}
\end{document}